\documentclass[aps, reprint, superscriptaddress, nofootinbib]{revtex4-1}
\pdfoutput=1
\usepackage{amsmath, latexsym, amssymb, hyperref, graphicx, color, slashed,xcolor}
\definecolor{nicered}{rgb}{0.7,0.1,0.1}
\definecolor{nicegreen}{rgb}{0.1,0.5,0.1}
\hypersetup{colorlinks, citecolor=nicegreen,linkcolor=nicered}
\begin{document}

\title{Strong CP violation: problem or blessing?}

\author{Goran Senjanovi\'{c} }
 \affiliation{International Centre for Theoretical Physics, Trieste, Italy}
\affiliation{Tsung-Dao Lee Institute \& Department of Physics and Astronomy, SKLPPC, Shanghai Jiao Tong
University, 800 Dongchuan Rd., Minhang, Shanghai 200240, China}

\author{ Vladimir Tello }
 \affiliation{International Centre for Theoretical Physics, Trieste, Italy}

\date{\today}

\begin{abstract}
        
  We readdress the issue of strong CP violation both in the Standard Model and in the Minimal Left-Right Symmetric Model and try to clear the confusion that seems to still pervade the field. We argue that the smallness  of strong CP violation, while harmless and basically decoupled from the rest of physics in the SM, in the context of LR symmetry provides a blessing by helping to narrow down the parameter space of the theory and connecting apparently uncorrelated physical quantities. In particular, in the context of left-right symmetry being parity, it either points to the suppression of leptonic CP violation noticed before, or it leads to relatively light right-handed neutrinos, potentially accessible at the next hadron collider. The latter, more natural in view of complex quark Yukawa couplings, goes hand in hand with the smallness of lepton flavor violation and enhances the possibility of observing  neutrinoless double beta decay.

    \end{abstract}

\maketitle

%
%  Introduction
%
\section{Strong CP violation in the Standard Model} 

Strong CP violation provides a beautiful interplay between strong and electro-weak interactions. Its measure manifestly illustrates this point 
\begin{equation}\label{thetabar}
\bar \theta = \theta  + \text{arg}\, \text{det} (M_u M_d),
\end{equation}   
where in obvious notation we have up and down quark mass matrices and $\theta$ is a vacuum selection parameter of the QCD defined by ($a, b = 1,...8$ is the color index) 
\begin{equation}\label{theta}
L_{QCD} = - \frac{1}{4} G_{\mu \nu}^a G^{\mu \nu, a} + \frac{\theta}{ 16 \pi^2} \epsilon^{\mu \nu \alpha \beta} G_{\mu \nu}^a G_{\alpha \beta}^a.
\end{equation}   
  Although we have separated the two contributions, from $\theta$ and quark mass matrices, in the Standard Model the physical parameter is only $\bar \theta$, since neither $\theta$ nor quark mass matrices can be measured on their own and the axial anomaly connects the two. Furthermore, it is important to stress that $\theta$ is not an ordinary coupling, but rather a boundary condition since $\epsilon^{\mu \nu \alpha \beta} G_{\mu \nu}^a G_{\alpha \beta}^a$ is a surface term.
  
  The separation becomes useful  in a theory beyond the Standard Model that can address the issue of the hermiticity of quark mass matrices. An example of such a theory is the Left-Right Symmetric Model of the spontaneous breakdown of parity in weak interactions, the principal topic of our work. In order to set the stage, and prepare the reader for a relatively painless discussion of our conclusions, we need to critically readdress the question of strong CP violation in the SM. Namely, it is generally considered to be the problem - and if this was true, one would conclude the same for the LR symmetric model. We wish to convince the reader that the prevailing wisdom is not correct, and that in reality this so-called problem can be a blessing in disguise for a theory beyond the SM - although strictly speaking in the SM it is neither of the two. 
  
  We are rushing ahead of ourselves though, first things first. 
   The experimental limit from the electric dipole moment of the neutron implies a strong constraint~\cite{Baluni:1978rf}
\begin{equation}\label{thetabarlimit}
\bar \theta \lesssim 10^{-10},
\end{equation}   
with a similar limit arising form mercury atoms~\cite{Graner:2016ses}. 
This has led to the proclamation of the strong CP problem. It is a strange proclamation for there is nothing that tells us what the value of $\bar \theta$ should be. It does not get renormalised, being a non-perturbative phenomenon, and is thus self-protected. In this sense, it is similar to the small electron Yukawa coupling, although unlike in the latter case there is no protective symmetry. 
%This said, the issue of the smallness of the strong CP violation is surely of a profound physical interest and we are by no means arguing against its importance.
% It is not a problem, though, since there is nothing inconsistent about it in being small in the SM.

The argument often does not stop here though. Sometimes, even in the context of the SM, one separates the two contributions in \eqref{thetabar} and talks about the apparent weak-interaction part  $\text{arg}\, \text{det} (M_u M_d)$. The argument then goes as follows.
We should expect this parameter to be much larger due to the weak interaction CP violation in the KM mixing matrix, and worse,  its smallness will not be respected in perturbation theory - hence the problem of naturalness. This has been repeated often since its first conception more than forty years ago, in spite of being wrong. As shown in~\cite{Ellis:1978hq}, the radiatively induced finite value of $\text{arg}\, \text{det} (M_u M_d)$  appears at the three-loop level and is vanishingly small, some nine orders of magnitude bellow the experimental limit. 

This should not come a surprise at all, it could be guessed (with some education) without doing the computation. First of all, the weak CP violation invariant parameter is not just the phase $\delta$, rather the tiny product $ \theta_1 \theta_2 \theta_3 \delta \simeq 10^{-5} $, where $\theta_i$ are the CKM mixing angles. A quick glance suffices to see the further loop and quark mass suppression - it should have been obvious that there could not have been a strong CP problem in the SM. The $\text{arg}\, \text{det} (M_u M_d)$ is structurally protected in the SM to an extraordinary degree. 

Although reassuring, strictly speaking this has no physical meaning  since $\text{arg}\, \text{det} (M_u M_d)$ is not a physical parameter in the SM. 
 All one can do is to compute the pure electro-weak part of the electric dipole of nuclei, which in the SM are extremely small, allowing, at least in principle, to measure the effect of $\bar \theta$ in the future~\cite{Chupp:2017rkp}. In any case, the real task, whether in the SM or any theory beyond, is to untangle the weak and strong contributions to the electric dipole moments.

We can conclude - ironically - that the only problem with the strong CP violation in the SM is that it is not a problem at all, for it offers no help in further reducing the parameter space of the theory. More precisely, the strong CP violation does not constrain any of the physical parameters of the SM, besides of course $\bar \theta$ itself. Although this is recognised by the experts, it is often said that one would like to know the tree-level value of $\bar \theta$, or at least why it is so small, something that the SM cannot address.
Strangely enough, one equally often calls this issue a problem - and when you argue that there is a problem, you must look for a solution. 

Before we do that, we wish to emphasise that nothing we said is new or original, but we believe that it was crucial to put this issue in the proper light in order to understand better its implications. If at this point we succeeded in having the reader share our outlook on the strong CP violation in the SM, she is likely to accept our findings in what follows. If not, if she persists in the belief that the problem is genuine, and that one must 'explain' why $\bar \theta$ is so small, our conclusions will definitely appear as unjustifiably euphoric. We continue though in the spirit that there is nothing to explain, the smallness of strong CP violation is just another of small numbers that plays a useful role in phenomenology, just the way the smallness of flavour violation in neutral currents leads to small charm quark mass.

\subsection*{Strong CP: blessing for the BSM} \label{beyondsm}

The so-called ``solutions'' to this non-existing problem can be divided in two categories: (i) attribute parity $\mathcal{P}$~\cite{Beg:1978mt} and/or CP violation~\cite{Georgi:1978xz} to spontaneous breaking and (ii) promote $\bar \theta$ to a dynamical variable as in the case of the Peccei-Quinn mechanism (PQ)~\cite{Peccei:1977ur}. In both cases, however, one trades the smallness of the tree-level $\bar \theta$ of the SM for the smallness of parameters that break the symmetries assumed in the above approaches. In the latter case one has to assume, ad-hoc, that the Yukawa couplings which violate the PQ symmetry must be ridiculously small - and this is surely far less natural than the smallness of $\bar \theta$ itself. 
In the former case, similarly, one must believe in basically exact discrete symmetries. But then, what is wrong with simply proclaiming that  $\bar \theta$ is small in the SM? 
In this sense, basically all the proposed models based on these ideas fare no better than the SM in explaining the smallness of strong CP violation. 
 
 This said, they may offer something new. Although one may not know why $\bar \theta$ is so small, one may correlate it to other physical quantities. In other words, the strong CP violation can becomes dynamical. This is especially true for the PQ mechanism that leads to the existence of the axion~\cite{Weinberg:1977ma}, among others a possible source of dark matter.
 What the axion field does is to provide the screening of $\bar \theta$, but once again cannot say anything {\it per se} about the the value of $\bar \theta$ or better, the value of the electric dipole moment of nuclei (or atoms).
 
  In the discrete symmetry approach one can obtain useful constraints on the parameter space of a model in question, unlike in the SM. Of particular interest is the left-right symmetric theory~\cite{LR}, put forward in order to have a dynamical theory of parity violation in weak interactions. This theory led to the existence to the right-handed (RH) neutrino and neutrino mass, long before it was suggested by experiment. In recent years, the modern version of this theory, the Minimal Left-Right Symmetric Model (MLRSM)~\cite{Minkowski} has been shown to be a self-contained, predictive theory of neutrino mass~\cite{Nemevsek:2012iq, Senjanovic:2019moe}, potentially accessible at the LHC and it is thus of imperative importance to study the consequences of strong CP violation in this theory. The situation depends critically whether one uses (generalised) parity  $\mathcal{P}$ or (generalised) charge conjugation  $\mathcal{C}$ as the LR symmetry, and must be treated separately. 
 
 The case of  $\mathcal{P}$ was discussed at length in~\cite{Maiezza:2014ala} but we believe it is still worth revisiting it in order to have a clear grasp of the phenomenological implications. Just like in the PQ approach, in this case one may tempted to speak of having a theory of strong CP violation. In the PQ case, $\bar \theta$ variable becomes dynamical, the vacuum expectation value of the axion filed - up to the explicit PQ symmetry breaking terms that must (and can) be assumed small. In the MLRSM it is parity breaking that is rendered dynamical - again up to the explicit  $\mathcal{P}$ breaking terms that must (and can) be assumed small, and this allows a unique opportunity to separate the $\theta$ and $\text{arg}\, \text{det} (M_u M_d)$ contributions. 

Not surprisingly then, certain CP violating parameters end up being tiny. Their stability requires either real neutrino Dirac Yukawa couplings (hard to keep without protective symmetry)~\cite{Kuchimanchi:2014ota} - or the right-handed neutrinos have to be somewhat light compared to the LR breaking scale. In the latter case, no fine-tuning is needed whatsoever since the obtained limits are technically natural. Ironically - contrary from what is usually preached - strong CP violation helps in reducing the parameter space of the theory. Just as in the SM, it is not a problem at all, but unlike in the SM case, it plays a useful role in pinning down the theory. 

 This leads to important correlations with other physical process, such as lepton number and lepton flavour violation. We should stress that the main contribution of our paper is to point out these important physical connections which shed light on the strong CP violation in the MLRSM. The beauty of the strong CP issue lies in the fact that one's conclusions depend crucial on the interpretation of the question and thus require great conceptual care, which we believe is equally important as the technical aspect itself. We have basically nothing to add to the latter, and in all honesty, not so much even to the former aspect, and an expert may be disappointed by what follows - still, we believe that our discussion may help the reader get a clearer picture of this fundamental issue. 
 
 On the other hand, the case  $\mathcal{C}$ is equivalent to the SM situation where one can say nothing about the strong CP violation {\it per se}, simply because the only physical parameter is $\bar \theta$, non-perturbative and thus insensitive to weak corrections. Just as in the SM, all one can do is to compute the weak part of different electric dipole moments. We discuss this in the central section \ref{section:strongCP} of this paper, where both cases are addressed.  
  
      Before we ask the reader to follow our discussion, though,
   in the next section we first review the salient features of the MLRSM in order to ease her pain. Our conclusions  are then left for the section \ref{section:outlook}, but it is worth emphasising once again that we are not after explaining why the strong CP violation is so small for it would end up being a tautology as in the most self-proclaimed ``explanations''. Instead, we simply wish to correlate this beautiful issue with other physical phenomena, which we believe is what physics is all about. This is the same attitude that~\cite{Maiezza:2014ala} took correctly, but the confusion persists when it comes to issues of naturalness and our aim is to once for all clear this up, at least in the context of the MLRSM, and to show that the resulting constraints are a blessing.

%
%In the  MLRSM, the seesaw mechanism follows naturally from spontaneous symmetry breaking, with $N$ mass proportional to the mass of the right-handed charged gauge boson, 
%$M_N\propto M_{W_R}$. The smallness of neutrino mass is thus linked to the near maximality of parity violation in weak interactions \cite{Mohapatra:1979ia} - in the limit of infinite $M_{W_R}$ one recovers the massless neutrino of the SM. 
%

%\vspace{0.1cm}
%{\bf Restored parity: probing neutrino mass.}\label{pand nu} The case of  $\mathcal{P}$ is however more involved. The Hermitian Dirac Yukawa couplings do not imply Hermitian mass matrices due to the complex vacuum expectation values in general. 

 %

\section{Minimal Left-Right Symmetric Model}

The MLRSM is based on the following gauge symmetry group 
\begin{equation}\label{LRgroup}
              \mathcal{G}_{LR} = SU(2)_L \times SU(2)_R \times U(1)_{B-L}                                                  
              \end{equation}
augmented by a discrete LR  symmetry which plays the role of left-right symmetry by interchanging left and right $SU(2)$ groups, and can be either $\mathcal{P}$ or $\mathcal{C}$. We discuss only the central features of the MLRSM, a reader in need of a more in-depth review of the theory would benefit from~\cite{Tello:2012qda}.

The fermion sector is given by the LH and RH  doublets
\begin{equation}
  \ell_{L,R} = \left( \begin{array}{c} \nu \\ e \end{array}\right)_{L,R}, \,\,\,\, q_{L,R} = \left( \begin{array}{c} u \\ d \end{array}\right)_{L,R}.                              
         \end{equation}
which transform under $\mathcal{P}$ as
\begin{equation}\label{underP}
             \ell_L \leftrightarrow \ell_R, \,\,\,\,\,  q_L \leftrightarrow q_R,                                  
\end{equation}
while under  $\mathcal{C}$ one has
\begin{equation}\label{underC}
             \ell_L \leftrightarrow \ell_R^*, \,\,\,\,\,  q_L \leftrightarrow q_R^*,                                  
\end{equation}

The original $G_{LR}$ symmetry is broken down to the SM gauge symmetry at the large scale $v_R$  by utilizing left and right $SU(2)$ triplets
  $ \Delta_L (3,1,2)$ and $\Delta_R (1,3,2)$,                
where the numbers in brackets denote the representation content under \eqref{LRgroup}. These scalars have the following form    
\begin{equation}
\Delta_{L,R}=\left(
 \begin{array}{c c}
%\frac{ \Delta_{L,R}^+ }{\sqrt{2}}
  \delta_{L,R}^+ /\sqrt{2}& \delta_{L,R}^{++} \\ [3pt]
\delta_{L,R}^0 & - \delta_{L,R}^+ /\sqrt{2}
%\frac{ \Delta_{L,R}^+ }{\sqrt{2}}
\end{array} 
\right)
\end{equation}
The essential point is the spontaneous breaking of LR symmetry achieved through  $v_L=\langle \delta_L^0\rangle= 0, v_R=\langle \delta_R^0\rangle\neq 0$. The vev $v_R$  is responsible for the masses of all non-SM particles, in particular the heavy gauge bosons $W_R$ and $Z_R$, and the RH neutrinos $N$, leaving only the SM gauge symmetry.

Next, the SM gauge symmetry breaking is achieved by the $SU(2)_L \times SU(2)_R$ bi-doublet $\Phi (2,2,0)$, which contains two SM doublets 
\begin{equation}
\Phi= \left[\phi_1, i\sigma_2 \phi_2^*\right],\quad \phi_i= \left( \begin{array}{c} \phi_i^0 \\ \phi_i^- \end{array}\right),\quad i=1,2. 
\end{equation}
 with the most general vev  given by
\begin{equation}\label{Phivev}
\langle\Phi\rangle=v\, \text{diag} (\cos\beta,-\sin\beta e^{-ia})
\end{equation}

   The quark Yukawa couplings  take the following form 
\begin{equation}\label{eq:quarks&phi}
L_Y=  - \overline{q_{L}}\,\big(Y_{1}^q \Phi - Y_{2}^q \,  \sigma_2 \Phi^* \sigma_2)\, q_R+\text{h.c.},
\end{equation}
 while their leptonic analogs are given by
\begin{equation}\label{Ly}
\begin{split}
\mathcal{L}_Y= & \,-\overline{\ell_L}(Y_1^{\ell}\Phi-Y_2^{\ell} \sigma_2 {\Phi}^* \sigma_2) \ell_R \,\\
&-\frac{1}{2}\left(\ell_L^TY_L i\sigma_2\Delta_L\ell_L+ \ell_R^TY_R i\sigma_2\Delta_R\ell_R\right)+\text{h.c.}
\end{split}
\end{equation}

 Under parity, consistent with \eqref{underP}, one has 
\begin{equation}\label{parity}
 \Delta_L\leftrightarrow \Delta_R,\quad \Phi\rightarrow\Phi^{\dagger}
\end{equation}
which imply Hermitian Dirac type couplings of the bi-doublet (both for quarks and leptons) and the same left and right couplings for the 
Higgs triplets
\begin{equation}\label{Yparity}
Y_{1,2}=Y_{1,2}^{\dagger}, \quad Y_L=Y_R.
\end{equation}
Since the bi-doublet vev is complex in general, see \eqref{Phivev}, the quark mass matrices do not have to be hermitian.

An essential comment is called for. We do not claim that the above relations are exact, for it makes no sense to speak of exact global symmetries. After all, unlike in the case of gauge invariance, one can move continuously away from a point of symmetry without losing anything - a point of symmetry stops being special, period. All we mean is that the spontaneous breaking of left-right symmetry dominates over the explicit one, just as one must argue in the case of PQ symmetry that the explicit breaking is small. We should stress that the explicit breaking, even extremely tiny suppressed by the Planck scale~\cite{Rai:1992xw} plays an important role in providing a way out of the domain wall problem~\cite{Zeldovich:1974uw} associated with the spontaneous breaking of an exact discrete symmetry. True, the domain wall problem would be in principle also solved~\cite{Dvali:1995cc} by the non-restoration of parity at high temperature~\cite{Mohapatra:1979vr}, but there is no proof that this happens due to subtleties of higher order effects in the temperature expansion~\cite{Bimonte:1995sc}. The bottom line is that the small explicit breaking of parity is actually welcome, and it is the only guarantee that there is no domain wall problem unless inflation takes place below the scale of LR symmetry, not the case in the MLRSM we are discussing here (we would have to go beyond the minimal model, but doing that would open Pandora's box of possibilities and the predictions would be lost).

To be precise, we are then only asking that this theory be self-contained and complete when it comes to fermion masses - including the neutrino one - which determines the precision of the validity of \eqref{Yparity}. With this cleared, we can now proceed.

Under charge conjugation, consistent with \eqref{underC}, one has 
\begin{equation}\label{charge}
 \Delta_L\leftrightarrow \Delta_R^*,\quad \Phi\rightarrow\Phi^T
\end{equation}
so that  the Yukawa couplings of the bi-doublet are symmetric (both for quarks and leptons)  and the left and right triplet Yukawas are complex conjugates of each other
\begin{equation}\label{Ycharge}
Y_{1,2}=Y_{1,2}^T, \quad Y_L=Y_R^*.
\end{equation}
In this case, the quark mass matrices are symmetric and thus one has $V_R = V_L^*$ before rotating $V_L$ into the CKM form by getting rid of unphysical phases. In the physical basis, one would have 
\begin{equation}\label{VRcaseC}
V_R = K_u V_{CKM}^* K_d,
\end{equation}
where $K_{u,d}$ are diagonal matrices of phases.

The SM Higgs doublet $h$ and the new heavy scalar doublet $H$ are given by
\begin{equation}
h = c_\beta \phi_1 + e^{-ia} s_\beta \phi_2,  \quad   H = - e^{ia} s_\beta \phi_1 + c_\beta \phi_2,
\end{equation}
where $c_\beta \equiv \cos_\beta, s_\beta \equiv \sin_\beta$ hereafter.  The new doublet $H$ mediates tree level flavor violation in the K and B-meson sectors, which imply $m_H   \gtrsim 20 \text{ TeV}$~\cite{Bertolini:2014sua} and take it out of the LHC reach.

In the neutrino sector the situation is more involved since the RH neutrinos become very heavy.  It is convenient to work with their charge conjugate states $N_L= C \bar \nu_R^{T}$, which from \eqref{Ly} implies $M_N=v_R Y_R^*$.
This  leads to the seesaw expression~\cite{Minkowski, rest} for light neutrino masses
  \begin{equation}\label{seesaw}
M_{\nu}= -M_D^T\frac{1}{M_N}M_D
\end{equation}
where $M_D = Y_D \, v =  (c_{\beta} Y_1^{\ell} + e^{i a} s_{\beta} Y_2^{\ell}) \,v$. The physical meaning of this is profound since one can produce $N's$ at hadron colliders through the Drell-Yan production of $W_R$ and actually measure $M_N$. The golden channel is the so-called KS process~\cite{Keung:1983uu},  consisting of both same and opposite sign charged lepton pairs accompanied by jets, which probes directly the Majorana nature of RH neutrinos while observing directly both lepton number and lepton flavor violation~~\cite{Tello:2012qda}. Moreover, the original LR symmetry allows to disentangle  the seesaw by predicting $M_D$ from $M_\nu$ and $M_N$, as shown originally in~\cite{Nemevsek:2012iq}.
  
%  The neutrino mass matrix is given as a function of $M_D$ and $M_N$. 

%
%The crucial ingredient is the RH analog $V_R^q$ of the CKM matrix $V_L^q$ which is rather sensitive to $s_a t_{2 \beta}$. In recent years we had managed~\cite{Senjanovic:2014pva} to solve the long-standing problem of computing analytically $V_R^q$  which has troubled the MLRSM for some forty years. It turns that $V_R^q$ takes  a simple approximate  form
%\begin{equation}
%\label{eq:VR}
%(V_R^q)_{ij} \simeq (V_L^q)_{ij} - i s_a t_{2 \beta}  \frac{(V_L^q)_{ik} ( V_L^{q \dagger}m_uV_L^q)_{kj} }{m_{d_k}+m_{d_j}}  
%+O(\epsilon^2) 
%\end{equation}
% It can be shown that the left and right mixing angles are almost the same, and right-handed phases depend only on $V_L$ and  
%$s_a t_{2 \beta}$. Thus by measuring $V_R^q$ one can predict the amount of parity violation in the gauge interactions of quarks.  In particular the near equality of LH and RH quark mixing angles justifies the experimental limits on $W_R$ mass~\cite{Aaboud:2019wfg}. The knowledge of $V_R^q$ leads furthermore to precise predictions for low energy processes, see e.g.~\cite{Bertolini:2019out}.

A small digression. Since the Yukawa couplings are complex in general, $CP$ is broken explicitly. There is also induced CP violation through the bi-doublet vev, measured by the small parameter $s_a t_{2\beta}$, with $s_a t_{2\beta}\lesssim 2m_b/m_t$~\cite{Senjanovic:2014pva}. When it vanishes the LR symmetry in the quark mass sector is maintained, and quark mass matrices become hermitian in the case of parity. Thus $s_a t_{2\beta}$ will play a crucial role in controlling strong CP violation. In particular, as we will see, in the case of parity $s_a t_{2\beta} = 0$ is consistent with $\bar \theta = 0$. We are getting ahead of ourselves though, first things first.

%The parity conserving limit $s_a t_{2 \beta} = 0$, motivated by the smallness of strong CP violation, is particularly clean since then one has the exact equality of the LH and RH quark mixing matrices. 

%  
%The SM symmetry breaking through $\langle \Phi \rangle$ induces a tiny vev $v_L$ of 
% the left-handed triplet $\Delta_L$ with  a hierarchy of $SU(2)_L$ breaking
%$v_L\propto v^2 /v_R$~\cite{MohSenj81}. The naturally small $v_L$ is self-protected~\cite{MohSenj81} (for a recent discussion, see \cite{Maiezza:2016ybz}) and is a direct source of neutrino mass, coined type II seesaw~\cite{typeII,MohSenj81}. 

\section{Strong CP in MLRSM}
\label{section:strongCP}

   We are now ready to discuss the constraints that the strong CP violation imply for the MLRSM. Since the situation is quite different in the cases of parity and charge conjugation as LR symmetries, we discuss them separately. We need to clarify some issues though before we plunge into our discussion. More often than not, the smallness of physical parameters has been treated as a problem in the modern day high energy physics. This by itself however does not make sense. After all,  how do we probe the theories of natural phenomena?  Before facing experiment, the parameter space of a theory in question is usually too big and needs to be restricted, thus the smallness of some parameters should rather be taken as a gift. Take for example flavor conservation in neutral currents - it requires the charm quark to lie much below the electro-weak scale. The heavy top quark requires tiny mixings with the second and especially first generation and so on and so forth. 
   
    Now, the experiment has started to put constraints  on the MLRSM parameter space by requiring that the new scale $v_R$ be large. LHC has already set a rigorous limit from through the  di-jet decay of $W_R$,  $M_{W_R} \gtrsim 4 \text{TeV}$~\cite{Aad:2019hjw}, and the low energy constraints from K and B meson physics, as we said require $m_H \gtrsim 20 \text{TeV}$. For this reason, 
    not much can be learned directly about the complex Higgs potential couplings and the constraints from the strong CP violation  help a lot as we show below. Contrary to the prevailing wisdom, the smallness of strong CP violation is not a problem whatsoever.
   In any case,  we are used to small couplings in the SM - think about the electron mass. The only thing one could worry about is the technical naturalness of small parameters, i.e. whether it amounts to large fine tunings. No such problem emerges in the MLRSM as we will see. 

\subsection{Strong CP and $\mathcal{P}$ } \label{P}

Parity plays a special role in the discussion of the strong CP violation, for it allows - unlike in the SM or the PQ case - to speak separately about the individual terms in $\bar \theta$~\cite{Beg:1978mt}. It is simply required by consistency of the theory since the Dirac Yukawa couplings must be hermitian, see \eqref{Yparity} which then requires $\theta$ to vanish separately in the  $\mathcal{P}$ symmetry limit. The exact limit is extreme, as we argued before - all that is needed that the explicit deviations are smaller than the spontaneous breaking.  Of course, one could argue that 
$\theta$ is zero due to the imposed parity symmetry, but that would be misleading. 

After all, a small explicit breaking serves to get of rid of dangerous domain walls as we discussed above - and strictly speaking saying that $\theta$ is small due to  $\mathcal{P}$ is a tautology, just like saying say that proton is stable due to baryon number conservation. Even worse, $\theta$ is a vacuum selection parameter  - so why should vacuum respect the symmetry that we like to impose~\cite{gia}? Another way of arguing is that $\theta$ is not the usual dimensionless coupling, but rather a kind of initial or boundary condition, since it accompanies a surface term. In short, while setting couplings to zero by a global symmetry is in a sense a tautology (proton is stable because baryon number is conserved say), arguing that parity implies $\theta = 0$ is simply wrong. If you wish, $\theta$ is another source of the spontaneous breaking of parity.

The bottom line is that, even if  $\mathcal{P}$ is not exact to start with, its explicit breaking being small is perfectly consistent (just as it is consistent to have small explicit PQ symmetry breaking). And when it comes to domain walls it may be as small as $M_R^2/M_{Pl}^2$, where $M_R$ and $M_{Pl}$ and the LR symmetry breaking and the Planck scale respectively, which for the LHC (or the next hadron collider) reach $M_R \simeq 10\,\text{TeV}$ is far below the experimental limit on $\bar \theta$. All we need to make sure is that the effects of the spontaneous symmetry breaking of parity do induce an unacceptably large value of $\bar \theta$, or better to say, the electric dipole moment of neutron. As we show now, everything is under control, and the consistency of the smallness of strong CP violation actually leads to physically interesting correlations. Thus, unlike in the SM, the smallness of $\bar \theta$ provides useful constraints on the theory.

Now, prior to symmetry breaking quarks are massless and so clearly there is no strong CP violation. What happens after the symmetry breaking?  
 The physical parameter is $\bar \theta$, so besides the value of $\theta$, it is therefore controlled  by $\text{arg}\, \text{det} (M_u M_d)$, which is turn determined by a small parameter $s_a t_{2\beta}$ that measures $\mathcal{P}$ and CP breaking in the quark sector. A careful study was made in~\cite{Maiezza:2014ala} who find (assuming no fine-tuning between $\theta$ and $\text{arg}\, \text{det} (M_u M_d)$)
\begin{equation}\label{thetabarP}
\bar \theta \simeq s_a t_{2\beta} \frac{m_t}{m_b},
\end{equation}   
 implying a practically vanishing $s_a t_{2\beta}$ (assuming of course that we do not fine-tune it against $\theta$, but that would be against the very idea of LR symmetry). This has important phenomenological implications and forces the new scale to be large, $M_{W_R} \gtrsim 20 \text{TeV}$, pushing  $W_R$ out of the LHC reach. 
 
 The question is whether this is stable in perturbation theory due to the $\mathcal{P}$ and CP breaking. It is easy to see that in the quark sector the answer is positive since small $s_a t_{2\beta}$ guarantees near hermiticity of quark mass matrices, and parity breaking comes only from $v_R$ that enters sub-dominantly at higher loops. Moreover, the KM CP violation enters also at high orders and is negligible as we argued in the Introduction. 
 
It was noticed by Kuchimanchi~\cite{Kuchimanchi:2014ota}, however, that leptonic CP violation affects the above conclusions. The argument goes as following. The physical measure  of strong CP violation, the $s_a t_{2\beta}$ parameter, is directly computable from the CP violating term in the Higgs potential (see e.g.~\cite{Maiezza:2016ybz} for a detailed discussion)  
\begin{equation}\label{epsilon}
 s_a t_{2\beta} = - 4 \frac {\text{Im}\,\alpha_2}{\alpha_3},
\end{equation}   
 where the $\alpha_{2,3}$ couplings correspond to the following terms of the Higgs potential
 \begin{equation}\label{alphas}
 \big[\alpha_2 \text{Tr}(\tilde{\Phi}\Phi^{\dagger}) +\text{ h.c.}\big]  \text{Tr}(\Delta_R\Delta_R^{\dagger})
	+ \alpha_{3} \text{Tr}(\Phi^{\dagger}\Phi\Delta_R\Delta_R^{\dagger}).
 \end{equation}
 It should be stressed that  $\alpha_3$ cannot be small for $M_{W_R} \lesssim 20 \text{TeV}$, since it is responsible for the mass of the heavy scalar doublet $H$ which must be rather heavy,  $m_H   \gtrsim 20 \text{ TeV}$~\cite{Bertolini:2014sua} due to its direct mediation of flavor violation.
 
 It is not surprising then that leptons matter since $\Delta_R$ couples directly to the RH neutrino $N$. The imaginary part of $\alpha_2$ is turned on at the one-loop level, and by using  \eqref{thetabarP} and \eqref{epsilon} it gives~\cite{Kuchimanchi:2014ota}
%  %
\begin{equation}\label{thetaloopP}
\bar \theta_{loop}  \simeq \frac{1}{16 \pi^2} \frac{m_t}{m_b} \text{Tr} (Y_R^\dagger Y_R [Y_1^\ell, Y_2^\ell])\, \text{ln}\frac{M_{Pl}}{v_R}
\end{equation}   
which, barring fine-tuning with the tree level, implies an important bound from \eqref{thetabarlimit}
\begin{equation}\label{limitonyukawas}
\text{Im}\, \text{Tr} (Y_R^\dagger Y_R [Y_1^\ell, Y_2^\ell])   \simeq \text{Im}\, \text{Tr}(Y_R^\dagger Y_R [Y_D, Y_\ell]) \lesssim 10^{-11}
\end{equation}   
where $Y_D$ and $Y_\ell$ are neutrino Dirac and charged lepton Yukawa couplings, respectively.

%We agree with the former possibility, however we show below that in the latter rather natural case, no fine tuning is needed - on the contrary it helps the theory be consistent with experiment.
%In the former case the author goes on to suggest possible changes of the theory, such as adding a PQ symmetry of breaking CP spontaneously by adding vector-like fermions. Since the MLRSM is self-contained, predictive theory of neutrino mass, we believe that one should stick to it to the bitter end. 
 Let us inspect carefully what the above limit really implies.
Obviously, the leptonic Yukawa couplings could be quite real (which cries for symmetry protection) and alleviate the situation~\cite{Kuchimanchi:2014ota}. This would have an important consequence of suppressing the leptonic electric dipole moments, since the leading contribution is proportional to the imaginary part of the product of $W_L-W_R$ mixing $\xi_{LR} \simeq M_{W_L}^2/M_{W_R}^2 t_{2 \beta} (c_a + i s_a)$ and  $Y_D$~\cite{Nemevsek:2012iq}. Since the imaginary part of of $\xi_{LR}$ is negligible, being directly proportional to $\bar \theta$, real $Y_D$ implies tiny leptonic electric dipole moments. 

However, as we argue now, it turns out that the Dirac Yukawa couplings do not have to be real. This is a surprising and fundamental result, since even if the Dirac Yukawas were complex as their quark counterparts - a natural scenario - the situation would be actually welcome.  Namely, smaller $Y_D$ means smaller $m_N$, which helps explaining the smallness of lepton number violation, see e.g.~\cite{Tello:2012qda}. We can actually be more precise about it.
         
      The crucial point is that in the MLRSM the seesaw can be untangled. In particular for the case of the Hermitian Dirac mass matrix $M_D$, relevant here, $M_D$ - and thus in turn $Y_D$ - can be determined from the knowledge of $M_\nu$ and $M_N$~\cite{Senjanovic:2016vxw}. For illustration, without any loss of generality, 
assume equal left and right-handed leptonic mixing matrices $V_L = V_R$ which then gives~\cite{Senjanovic:2016vxw}
  \begin{equation}\label {MDirac}
Y_D=i\, \frac{g}{2 M_W} V_L \sqrt{m_{\nu}m_N}V_L^{\dagger},
\end{equation}
where $V_L$ is the PMNS mixing matrix.  

Since $Y_R = g M_N/M_{W_R}$ and using \eqref{MDirac}, and by taking the largest value $y_\ell = y_\tau$, one can rewrite \eqref{limitonyukawas} as an estimate 
  %
%  \begin{equation}\label {MNvsMR}
%M_N^5 M_\nu \lesssim 10^ {-20} M_W^2 M_{W_R}^4.
%\end{equation}
%\textcolor{blue}{
\begin{equation}\label {MNvsMR}
M_N^5 M_\nu \lesssim 10^ {-16} M_W^2 M_{W_R}^4
\end{equation}
  %\begin{equation}\label {MNvsMR}
%M_N^5 M_\nu \lesssim 4 \times 10^ {-20} M_W^4 M_{W_R}^4/m_\ell^2.
%\end{equation}
%}
%
where we ignore the details of leptonic mixings without any real loss of generality, since they are not small.
This implies the following limit on the mass of $N$ by taking $M_\nu \simeq 10^{-10} \, \text{GeV}$
 %
%  \begin{equation}\label {MNlimit}
%M_N \lesssim 10^ {-6/5} \left(M_{W_R}/\text{GeV}\right)^{4/5} \,\text{GeV}.
%\end{equation}
%\textcolor{blue}{
  \begin{equation}\label {MNlimit}
M_N \lesssim 5 \times 10^ {-6/5} \left(M_{W_R}/\text{GeV}\right)^{4/5} \,\text{GeV}.
\end{equation}
%}
%
Thus, the RH neutrinos must be relatively light compared to $W_R$, just like most charged fermions of the SM compared to $W_L$. This fits nicely with their  possible impact on neutrinoless double beta decay and the limits from lepton number violation. The lower $v_R$, the smaller then $M_N$ and in turn $M_D$, making it easier to account for the smallness of strong CP violation. Of course, this is not very strict since the imaginary part of Dirac Yukawas could be somewhat small. Increasing the scale $v_R$ poses no real problem, all that is needed smaller $Y_R$ in order to comply with the strong CP violation limits.
% In order words,  strong CP violation points towards low scale LR symmetry.
 
  To get a feel for the scales, let us exemplify \eqref{MNlimit} with $M_{W_R} \simeq 10 \, \text{TeV}$ which gives 
  \begin{equation}
  M_N \lesssim  500\, \text{GeV} 
  \end{equation} 
  in the right ballpark for the observation of the KS process. 
 
  How natural are the above constraints, in a technical sense of the word? Since fermion masses are protected by chiral symmetries, there is absolutely no problem  and since most of the SM Yukawas are small, there is also nothing unnatural in an intuitive sense. 
  
%  After all, charmed quark must be light in order to suppress neutral current flavor violation, in complete analogy with the above bound on RH neutrino masses. Or think of a heavy top quark that forces the 1-3 mixing angle to be very small. Good theories live dangerously.
  
 It may happen that the $N$'s are too light to be seen at colliders. In particular if the lightest $N$ is the warm dark matter~\cite{Dodelson:1993je}. Still in this case, $M_N$ ends up being completely determined~\cite{Nemevsek:2012cd} with $m_N \simeq \text{KeV} - \text{GeV}$, making Dirac Yukawa couplings much smaller, so that the strong CP violation bound is automatically satisfied.
 
    Needless to say, there is always the contribution due to the KM phase, but that is negligible as in the SM.
Thus, in a nut-shell, the smallness of $\bar \theta$ implies either a real Dirac Yukawa matrix, which in turn implies small leptonic EDM or, more naturally, small $N$ masses as we have shown. In either case, the smallness of strong CP violation makes the theory more predictive.
 
 \subsection{Strong CP and C: connection lost } \label{C}
 
 If the LR symmetry was charge conjugation $\mathcal{C}$, the situation would be less restrictive since at the tree level $\bar \theta$ is undetermined, just as in the SM. Namely, in this case the quark mass matrices are symmetric, the $\text{arg} \, \text{det} (M_u M_d)$ is arbitrary and so is $\theta$ itself  since it does not break $\mathcal{C}$. The situation is similar to the SM one, the only physical parameter is $\bar \theta$, and, unlike in the  $\mathcal{P}$ case, one has no right to separate the strong and weak contributions. In other words, it makes no sense to speak of  radiative corrections to $\bar \theta$, as it makes no sense to do it either in the SM.
 
   One could still pretend to study the loop contribution to  $\text{arg} \, \text{det} (M_u M_d)$, as done in~\cite{Ellis:1978hq}.
   We know that in the SM these corrections are negligible, so we would have to bring in the physics of RH gauge bosons and new scalars into the game. It is easy to see that the dominant contribution would appear at the one-loop level through $W_L - W_R$ mixing as in Fig. 1.
%   
%   with an estimate
% %
%  \begin{equation}\label {thetabar}
%\text{arg} \, \text{det} (M_u M_d) (loop) \simeq \frac{\alpha}{4 \pi}\, \text{Im}\,  \text{Tr}  (\xi_{LR} K_d^* V_{CKM}^T K_u^* \frac {m_u}{m_d} V_{CKM}) 
%\end{equation}
%%
% where we used \eqref{VRcaseC} and $m_{u,d}$ stand for diagonal up and down quark mass matrices. In the physical basis with real and positive quark masses all the complexity is gone into the left and right mixing matrices. 
 
 One could go on and put constraints on the three contributions in the above equation, coming from the  KM phase, the $K_{u,d}$ phases and the complex $W_L - W_R$ mixing. It would be wrong though. Once again, $\text{arg} \, \text{det} (M_u M_d)$ is not a physical parameter and it cannot be computed, period. Instead, one could use the Fig. 1 as a typical contribution to the electric dipole of a quark, but it is subdominant contribution to the EDM of the neutron. Namely, it is suppressed by a small light quark mass which goes away with the strong interactions.  The leading contribution comes the chiral loops, and for recent discussions in the context of the MLRSM see e.g.~\cite{Maiezza:2014ala, Bertolini:2019out}.  Simply on dimensional grounds one expects for the EDM of the neutron
  \begin{equation}\label {edmC}
  d_n \simeq e \frac{1}{16 \pi^2} \,m_{\mathcal{N}} \, G_F \, \text{Im} \,V_{L ud} V^*_{R ud}\, \xi_{LR}
\end{equation} 
 where $\xi_{LR} \simeq M_{W_L}^2/M_{W_R}^2 t_{2 \beta} (c_a + i s_a)$ and $m_{\mathcal N} \simeq \text{GeV}$ is the nucleon mass,  so that $m_{\mathcal{N}}G_F/16\pi^2 \simeq 10 ^{-21} \text{cm}$. Then, experimental limits on the neutron EDM, $d_n\lesssim 10^{-26} \,e\cdot \text{cm}$, imply
  %
  %\begin{equation}\label {edmC}
  %d_n \simeq 10 ^{-20} \,\text{cm} \, \text{Im} \,V_{L ud} V^*_{R ud} \,\xi_{LR}
%\end{equation} 
%\textcolor{blue}{
 \begin{equation}\label {edmC}
%  d_n \simeq 10 ^{-21} \,e\cdot \text{cm} \, \times
  \text{Im} \,V_{L ud} V^*_{R ud} \,\xi_{LR}
  \lesssim 10^{-5}.
\end{equation} 
%}
% 
In view of the smallness of the mixing $\xi_{LR}$ there is basically no limit on the $V_R$ elements. The difference between the $\mathcal{P}$ and $\mathcal{C}$ case is really striking. 

 % Using $V_R = K_u V_L^* K_d$, and the lower limit  $M_{W_R} \gtrsim 4 \text{TeV}$, this implies limits for the $W_R$ accessible at the LHC
   %
  %   \begin{equation}\label {limitsC}
  %c_a t_{2\beta}\text{Im}\, K_{u}^{11}K_{d}^{11} \lesssim 10 ^{-1}\end{equation} 
%
 % \begin{equation}\label {limitsC}
  %\text{Im} K_{\,u,d}^{11} \lesssim 10 ^{-2} \,\,\,\,\,\,\,\, s_a t_{2 \beta} \lesssim 10 ^{-2}.
%\end{equation} 
% 
 % As the $W_R$ mass gets increased, these limits get weaker.
 %This constraint, albeit not very strong, is important since it affects the low energy limits on the LR breaking scale. 
% 
% There is a similar term with $m_u$ and $m_d$ reversed, but is subdominant. Clearly, the leading term will involve $m_t/m_b$ due to the small quark mixings, just as in the $P$ case. It gives roughly (we ignore small terms such as $\text{ln} M_{W_R}/M_{W_L}$
%  %
%  \begin{equation}\label {thetabarCfinal}
%\bar \theta (loop) \simeq \frac{\alpha}{4 \pi}\, \frac {m_t}{m_b}\, \text{Im}\, (\xi_{LR} V_{CKM, 3 i}^2 K_{d, ii}^*  K_{u, 33}^*).
%\end{equation}
%
% 

\begin{figure} \begin{center} \includegraphics[height=3.5cm]{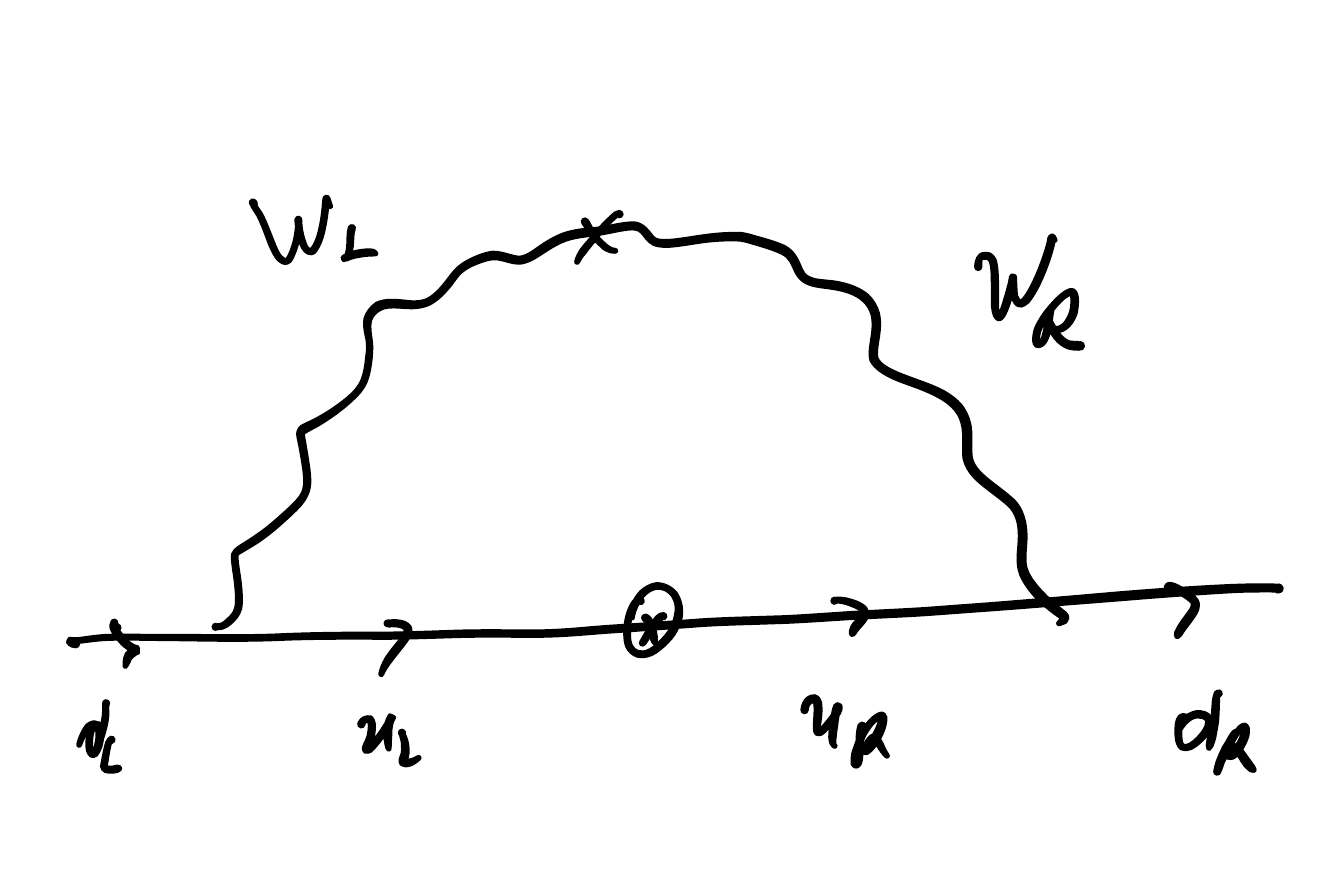} \caption{Typical diagram for the one loop contribution to $\bar \theta$ in case $\mathcal{C}$. As we argue in the text, this has has no physical meaning whatsoever - one must and can compute only the direct electric dipole moments.} \end{center} \end{figure}

%There are three contributions in the above equation, coming from the  KM phase, the $K_{u,d}$ phases and the complex $W_L - W_R$ mixing. The former is automatically small since the KM phase enters through small quark mixing angles. 
%The situation now depends on the size of the  $W_L - W_R$ mixing $\xi_{LR} $ which depends in turn of the mass of $W_R$ and the $s_a t_{2 \beta}$ parameter
%  %
%  \begin{equation}\label {LRmixing}
% \xi_{LR} \simeq \frac{M_{W_L}^2}{M_{W_R}^2} t_{2 \beta} (c_a +  i s_a)
%\end{equation}
%%
%and thus goes to zero in the limit of infinitely large LR breaking scale. From the experimental limit
%$M_{W_L}^2/M_{W_R}^2 \lesssim 10^{-3}$, the smallness of $\bar \theta$ implies $s_a t_{2 \beta} \lesssim 10^{-6}$, four orders of magnitude bigger than the limit in the case of $P$. Similarly, one has $ \text{Im} (K_u K_d) \lesssim 10^{-6}$. The new phases are practically negligible and one predicts $V_R \simeq V_L = V_{CKM}$. Again, it is easy to convince oneself that, as in the case $P$, the small parameters are naturally small in the usual, technical sense.
%
%  
%
%

%
%\vspace {2cm}
% 
%This in turn leads to the neutrino mass matrix to the leading order in $M_D/M_N$
%  \begin{equation}\label{seesaw}
%M_{\nu}=\frac{v_L}{v_R} M_N^* -M_D^T\frac{1}{M_N}M_D
%\end{equation}
%
  
%   \subsection{Determining light and heavy neutrino mass matrices}\label{MnuMN}  

In short, independently of the details of the LR symmetry implemented, be it  $\mathcal{P}$ or $\mathcal{C}$, one cannot predict the value of $\bar \theta$ in the MLRSM, just as one cannot do it in the PQ case or basically any other theory, including the SM. In the case of  $\mathcal{P}$ one obtains useful constraints needed to keep $\bar \theta$ under control, but its value itself can be determined only from experiment. What is needed is to measure different electric dipole moments in order to untangle the weak and strong contributions~\cite{juan}.

\section{Summary and outlook}
\label{section:outlook}

We have revisited the issue of strong CP violation in the MLRSM with the conclusion that the smallness of $\bar \theta$ helps constrain a number of unknown parameters of the theory. There is nothing technically unnatural that takes place, which provides yet another example besides the SM, where there is no strong CP problem. All that one learns is that the spontaneous CP violating parameter $s_a t_{2 \beta}$ has to be tiny, implying a very small imaginary part of a particular quartic coupling in the Higgs potential. This we believe is a fundamental point which illustrates nicely the fact that strong CP violation is a not a problem {\it per se} at all. On the contrary, it may even be a benediction for a well defined self-contained theory such as the one in question. 

The situation depends critically on whether the LR symmetry is parity or charge conjugation. In the latter case, the 
question of $\bar \theta$ cannot be addressed perturbatively, just as in the SM - all one can do is to compute the various electric dipole moments. There have been studies before and we have stayed away from it here.

  In the former case, the situation is different. Having  $\mathcal{P}$ allows to address separately the weak and strong contributions to $\bar \theta$ and to treat it perturbatively. The bottom line is that the 
  smallness of $s_a t_{2 \beta}$ is  maintained in perturbation theory as long as the RH neutrinos are fairly light $m_N \lesssim 500 \,\text{GeV}$ for a LR breaking scale around $\text{TeV}$ possibly reachable at the next hadron collider (small $s_a t_{2 \beta}$ takes the MLRSM out of the LHC reach). This at the same time accounts automatically for the smallness of LFV and provides a possibility that the RH physics dominates the neutrinoless double beta decay over the usual contribution of LH neutrino Majorana mass.
As $v_R$ increases, this limit relaxes but still, the essential prediction is the Yukawa couplings of  $N's$, must be small. This of course is both technically and intuitively natural, the SM model is all about small Yukawa couplings of quarks and charged leptons. 

  It may appear that the constraints persist even for large LR breaking scale, in contradiction with the situation in the SM. After all, the limit on $s_a t_{2 \beta}$ is independent of $M_{W_R}$. The point though is that $s_a t_{2 \beta}$ appears only through the new physics of heavy $W_R$ and thus effectively disappears in the SM limit - in other words, its smallness stops having physical meaning. Everything is consistent as it should be.
  
  Alternatively, as noticed in~\cite{Kuchimanchi:2014ota}, there is a less natural possibility of real neutrino Dirac Yukawa couplings, which implies an important prediction of small leptonic EDM. In any case, the smallness of strong CP violation plays a gratifying role of reducing the parameter space and implying important phenomenological predictions.   
      
   The constraints in question play particularly important role for the MLRSM accessible at the LHC or next hadron collider energies, but why would that ever happen?
  The answer lies in a deep connection with a neutrinoless double beta decay whose observation may signal the contribution of new physics if neutrino mass is not sufficiently large to do the job. Or if for example one could measure the polarization of electrons and find out that they are right handed, in which case the SM neutrino contribution would be ruled out. In this case $W_R$ could not be too heavy~\cite{Nemevsek:2011aa}, making the case for its manifestation if not at the LHC, surely at the new collider. 
  
     We close with a remark regarding the choice of the LR symmetry in defining the MLRSM. The original choice was parity, a logical assumption in view of LH fermions becoming RH under parity. The trouble is that charge conjugation is also a possibility, maybe less intuitive, but nonetheless a LR symmetry since LH fermion become LH anti-fermions, or complex conjugates of RH fermions. An apparent advantage of the latter choice is the prediction of symmetric quark and lepton mass matrices, which predicts same LH and RH quark mixing angles, and the form of neutrino Dirac mass matrix and untangle the seesaw~\cite{Nemevsek:2012iq}. 
     
     Furthermore,  $\mathcal{C}$ becomes a gauge symmetry in the $SO(10)$ grand unified theory, which adds to its appeal. However,  the same $SO(10)$ theory predicts astronomically large LR breaking scale and thus this arguments loses its physical relevancy. (By arbitrarily adding many new particles, one can of course lower arbitrarily the intermediate LR scale, but this in turns implies giving un on all predictions - this should not be considered a grand unified theory anymore, or a theory at all, for that matter.) 
     
     It turns out that one can untangle the seesaw also in the case $\mathcal{P}$~\cite{Senjanovic:2019moe} and moreover, one can predict both the RH quark mixing angles and phases with a single unknown parameter~\cite{Senjanovic:2014pva}. The possibility of controlling strong CP violation with the emerging  correlations discussed in this work make a strong case for the choice of parity in defining the left-right symmetric theory, the way it all started. The crucial thing to remember is that we do not require parity to be exact, as long as its spontaneous breaking dominates over the explicit one.  The reason lies in the original motivation of the left-right symmetric theory of having the breakdown of parity originate dynamically, as is the case with gauge symmetries.  This is in complete analogy with the axion picture where one demands that the explicit breaking of PQ symmetry be smaller that the spontaneous one.

\vspace{0.6cm}

\subsection*{Acknowledgments}

    G.S. acknowledges ongoing debates with Gia Dvali on the question of strong CP violation over the years. He wishes to thank Xiao-Gang He and other members of the TD Lee institute in Shanghai for their warm hospitality in the course of this work. We have benefited from fruitful discussions with Kaladi Babu, Alessio Maiezza, Fabrizio Nesti and Juan Carlos Vasquez. We are grateful to Babu and Dvali for helping us improve the manuscript.

\end{document}